\documentstyle[aps,epsf,multicol,amsmath,amsfonts,graphicx,bbm]{revtex}

\begin{document}


\title{Lossless quantum coding in many-letter spaces}
\author{Kim J. Bostr\"om}
\address{
}
\author{}
\address{Institut f\"ur Physik, Universit\"at Potsdam, 14469 Potsdam, Germany}
\date{Version 2.3 beta / \today}

\maketitle

\begin{abstract}

Based on the concept of many-letter theory, an observable is defined measuring the raw quantum information content of single messages. A general characterization of quantum codes using the Kraus representation is given. Compression codes are defined by their property of decreasing the expected raw information content of a given message ensemble. Lossless quantum codes, in contrast to lossy codes, provide the retrieval of the original data with perfect fidelity. A general lossless coding scheme is given that translates between two quantum alphabets. It is shown that this scheme is never compressive. Furthermore, a lossless quantum coding scheme, analog to the classical Huffman scheme but different from the Braunstein scheme, is implemented, which provides optimal compression. 
Motivated by the concept of lossless quantum compression, an observable is defined that measures the core quantum information content of a particular message with respect to a given \emph{a priori} message ensemble. The average of this observable yields the von Neumann entropy.

\end{abstract}

\begin{multicols}{2}

\narrowtext

\section{Introduction}

In~\cite{Bostroem_mlconcepts}, the concept of a quantum information theory generalized to messages with components of variable length has been presented, here referred to as \emph{many-letter theory}. Based on this concept, 
an observable is defined measuring the raw quantum information content of single messages. A general characterization of coding schemes using the Kraus representation is given. Compression then means decreasing the expected raw information content of a given message ensemble. Apart from lossy schemes like the Schumacher coding scheme, which compresses quantum messages by neglecting ``unimportant'' information, a lossless coding scheme can be implemented, which not only ensures perfect fidelity in retrieving the original messages but also provides optimal compression. This coding scheme differs from the Braunstein scheme presented in \cite{Braunstein98}, mostly in that it is a perfectly lossless code and, since it exploits the features of many-letter space, it cannot be implemented in a standard block Hilbert space. Motivated by the concept of lossless compression, a quantum mechanical observable is defined that measures the amount of core quantum information contained in a particular message with respect to a given \emph{a priori} message ensemble.

This paper is separated into two parts. The first part reviews roughly the basic concepts of classical coding in order to motivate the corresponding notions presented in the second part, which is dedicated to quantum coding. A detailed summary of classical information theory can be found in~\cite{MacKay}, a very recommendable review on quantum information theory is given in~\cite{Preskill}.

\section{Classical information theory}

\subsection{Notions and definitions}

The following notions and definitions are used throughout this paper. The reader is referred to~\cite{Bostroem_mlconcepts} for further details.

A classical message is a string $\boldsymbol x$ of letters $x$ taken from an alphabet ${\cal A}$ of size $|{\cal A}|$ and is denoted by
$\boldsymbol x=(x_1\cdots x_n)$. Strings of length $n$ are explicitely denoted by
\begin{equation}
	x^n:=(x_1\cdots x_n)\quad.
\end{equation}
The set of \emph{block messages} $x^N$ of fixed length $N$ is written as
\begin{equation}
	{\cal A}^N:=\{(x_1\cdots x_N)\mid x_n\in{\cal A}\}\quad.
\end{equation}
Let us also allow for the \emph{empty message} $x^0=(\cdot)$ that forms the set
${\cal A}^0:=\{(\cdot)\}$.
The set of all messages of finite length is defined by
\begin{equation}
	{\cal A}^+:=\bigcup_{n=0}^\infty{\cal A}^n\quad.
\end{equation}
A general message ensemble is represented by a random variable
\begin{equation}
	\boldsymbol X:=\{[x^n,p(x^n)]\mid x^n\in
	\Omega\}\quad,
\end{equation}
of strings $x^n$ drawn with \emph{a priori} probabilities $p(x^n)>0$ from a source set $\Omega$, such that $\sum_{x^n\in\Omega}p(x^n)=1$.
A \emph{canonical message} $x^N$ is drawn from the ensemble
\begin{equation}\label{canonical}
	X^N=\{[x^N,p(x^N)]\mid x^N\in{\cal A}^N\}
\end{equation}
with factorizing \emph{a priori} probability $p(x^N)=p(x_1)\cdots p(x_N)$.

\subsection{Raw Information content}\label{inf_cont}

There are several ways to think about an ``amount of information'' carried by messages. I will take a rather pragmatic point of view by simply asking ``How much effort does it take to communicate the message?''. Say Alice builds a device to send messages to Bob, who in turn builds the adequate receiver. To every single letter Alice has to use a sender unit that creates any of the alphabet letters. The more letters in the alphabet, the more complicated the device, hence the more effort to communicate the messages. If the length of the message is $N$, then $N$ of the sender units are in use. Bob builds enough receiver units to receive a message of arbitrary length. He adds a meter on top of his receiver indicating the number of active receiver units and calls the indicated value the \emph{raw information content} of the received message. He calibrates the pointer to show exactly 1 unit of information whenever the message contains the smallest amount of information, given by a single-letter message over the binary alphabet. He calls the unit of this information ``1 bit''. To put it mathematically, we define the \emph{raw information content} as a function $I:{\cal A}^+\rightarrow[0,\infty)$ with
\begin{equation}\label{single_inf}
	I(\boldsymbol x):= \log|{\cal A}|\,L(\boldsymbol x)\quad,
\end{equation}
where $L(\boldsymbol x)$ is the length function on ${\cal A}^+$. The unit of this information measure is ``1 bit''.
For example, the iformation content of a message of length $n$ over the binary alphabet ${\cal A}=\{0,1\}$ is $I(x^n)=\log 2\,L(x^n)=n$ bits.
Note that this measure applies to single messages rather that message ensembles. No statistical information is needed, it is just an observable that can be realized by some measuring apparatus. The value of $I$ is indicated by a disk manager as the file size on a hard disk or by an internet browser as the received information  during a download process. The bigger this number, the longer it takes to open, edit, save or download a particular message. In this sense it is a true physical observable.
Taking into account the statistical properties of a given message ensemble $\boldsymbol X$ one may define the \emph{ensemble raw information content} by
\begin{equation}\label{ensemble_inf}
	\overline I(\boldsymbol X)=
	\sum_{\boldsymbol x\in\Omega}p(\boldsymbol x)
	\log|{\cal A}|\,L(\boldsymbol x)\quad.
\end{equation}
The bigger this number, the bigger the effort of communication on the average.  
Although a lot of other information measures are possible, the measure (\ref{single_inf}) and its average~(\ref{ensemble_inf}) will suffice for our purposes.

\section{Classical coding}

\subsection{General types of codes}

Alice and Bob decide to use a code while exchanging their messages. If encoding and decoding is easy but the decoding scheme is hard to guess, it ensures the \emph{security} of their conversation and falls in the domain of \emph{cryptography}. If the code takes advantage from the statistical properties of the used message ensemble, it allows for \emph{compression} in order to minimize the effort of transmission or storage of the data. Mathematically, a code is just a mapping $c$ from a given source set $\Omega$ of messages composed from a source alphabet ${\cal A}$ to a code set $\Omega_C$ of code messages composed from a code alphabet ${\cal A}_C$. A code can be specified into two types:
\begin{itemize}
\item
	A \emph{lossless} code is uniquely decodeable, i.e.
	$\forall\boldsymbol x, \boldsymbol y\in\Omega, \boldsymbol x\neq\boldsymbol y: c(\boldsymbol x)\neq c(\boldsymbol y)$.
	For any finite set $M\subset\Omega$ we have $|M|=|c(M)|$.
\item
	A \emph{lossy} code maps certain messages to the same encoding, i.e. $\exists \boldsymbol x,\boldsymbol y\in\Omega: c(\boldsymbol x)=c(\boldsymbol y)$. For every finite set $M\subset\Omega$ we have $|M|\geq|c(M)|$.
	Each time a message is being irreversibly encoded, the decoder cannot recover the original message and will give an error. If the probability of error can be made very small, the lossy code may be useful.
\end{itemize}

Furthermore, there are two other important types of code:
\begin{itemize}
\item
	A \emph{block code} encodes only block messages of fixed length $N$ over a source alphabet ${\cal A}$ to block code messages of fixed size $M$ over a code alphabet ${\cal A}_C$. It is a function $c:\Omega\subset{\cal A}^N\rightarrow \Omega_C\subset{\cal A}_C^M$.
\item
	A \emph{symbol code} encodes messages of any length by encoding each letter separately. If $c:{\cal A}\rightarrow {\cal A}_C$ is a code on the source alphabet ${\cal A}$, then it can be extended to the code $c:{\cal A}^+\rightarrow{\cal A}_C^+$ by
	\begin{equation}\label{def_symbol}
		c(x_1\cdots x_n):=c(x_1)\cdots c(x_n)\quad.
	\end{equation}
\end{itemize}
A code $c$ thus maps a source message ensemble $\boldsymbol X$ to a code message ensemble $\boldsymbol Y=c(\boldsymbol X)$, which can be expressed in terms of the source message ensemble as
\begin{equation}
	\boldsymbol Y=\{[c(\boldsymbol x),p(\boldsymbol x)]\mid 
	\boldsymbol x\in\Omega\}\quad.
\end{equation}
The transformation to the new ensemble
\begin{equation}
	\boldsymbol Y=\{[\boldsymbol y,p_C(\boldsymbol y)]\mid \boldsymbol y\in\Omega_C\}\quad,
\end{equation}
is given by $\boldsymbol y:=c(\boldsymbol x)$, $\Omega_C:=c(\Omega)$ and
$p_C(\boldsymbol y):=\sum_{\boldsymbol x\in\Omega}p(\boldsymbol x)\,\delta(c(\boldsymbol x),\boldsymbol y)$, where
\begin{equation}
	\delta(\boldsymbol x,\boldsymbol y):=
	\begin{cases}
		1&;\,\boldsymbol x=\boldsymbol y\\
		0&;\,\boldsymbol x\neq\boldsymbol y
	\end{cases}
\end{equation}
is the string version of the Kronecker delta.	
Note that if $\boldsymbol X$ is a canonical message ensemble $X^N$, the code message ensemble $\boldsymbol Y=c(\boldsymbol X)$ is generally not.

\subsection{Binary symbol codes}

A binary symbol code is a symbol code $c:\Omega\subset{\cal A}^+\rightarrow\Omega_C\subset\{0,1\}^+$.
There is a connection between reversibility of binary symbol codes and the lengths of the encoded letters. It is given by the \emph{Kraft inequality} that states that the codeword lengths of a lossless binary code must satisfy
\begin{equation}
	\sum_{x\in{\cal A}} \left(\frac12\right)^{L_c(x)}\leq 1\quad,\label{kraft}
\end{equation}
where $L_c(x)$ is the length of the codeword corresponding to $x$.

\subsection{Prefix codes}

How can Bob decode a symbol code he received from Alice? By the original definition~(\ref{def_symbol}) the code is obtained by encoding each source letter separately and then concetenating the codewords to an entire string. If a code is lossless it is nevertheless possible to decode the message, since by construction there is a distinct code for any of the source messages. Among the lossless symbol codes there is an important class of code called \emph{prefix codes}. They are defined by the property that no codeword is a prefix of another codeword. Thus a prefix code is \emph{instantaneous}, i.e. it can be decoded simply from left to right without looking at the entire string. An example for prefix codes are telephone numbers. The decoder does not have to wait until the entire phone number is entered, it can proceed connecting while the numbers are sequencially transferred.
As soon as it arrives at a single telephone device, the connection is established. Luckily, one can prove that whenever the codelengths of a symbol code satisfy the Kraft inequality~(\ref{kraft}), there is a prefix code with the same codeword lengths (see \cite{MacKay} pp 95). In other words, whenever a given symbol code is lossless, it can be replaced by a prefix code with the same codeword lengths. Thus in the following we will always assume for a lossless symbol code to be a prefix code, so it can be instantaneously encoded and decoded.

\subsection{Compression codes}\label{comp_codes}

A code maps a given message ensemble to a code message ensemble with the codewords obtaining new lengths. Instead of using the length function $L_C:{\cal A}_C^+\rightarrow{\mathbbm N}$ on the code messages one can use a \emph{code length function} on the source messages, defined by
\begin{equation}
	\boldsymbol x\in{\cal A}^+:\quad 
	L_c(\boldsymbol x):=L_C(c(\boldsymbol x))\quad,
\end{equation} 
giving each source message $\boldsymbol x\in{\cal A}^+$ the length $L_c(\boldsymbol x)$ of its code. Consequently, each encoded message obtains an \emph{encoded information content},
\begin{equation}\label{code_single_inf}
	I_c(\boldsymbol x):= L_c(\boldsymbol x)\log|{\cal A}_C|\quad,
\end{equation}
and the encoded message ensemble obtains an \emph{encoded ensemble information content},
\begin{equation}\label{code_ensemble_inf}
	\overline I_c(\boldsymbol X)
	:=\sum_{\boldsymbol x\in\Omega}p(\boldsymbol x)I_c(\boldsymbol x)\quad.
\end{equation}
A \emph{compression code} is a code $c$ fulfilling
\begin{equation}\label{compression}
	\overline I_c(\boldsymbol X)\leq\overline I(\boldsymbol X)\quad.
\end{equation}

\subsection{Block compression}\label{block_comp}

Block compression (see e.g.~\cite{MacKay}, \cite{Preskill}, \cite{Shannon48}) applies to canonical messages $X^N$ of fixed block length $N$, generally given by~(\ref{canonical}). It is a lossy coding scheme in that it only encodes ``typical strings''. The clue of block compression is that the probability of error can be made arbitrarily small by increasing the block size $N$. 
Define the \emph{typical set} $T_\delta^N$ with \emph{tolerance} $\delta$ as the set of all messages $x^N$ drawn from the canonical ensemble $X^N$, whose \emph{a priori} probabilities fulfill
\begin{eqnarray}
	\quad 2^{-N(H+\delta)}&<&p(x^N)<2^{-N(H-\delta)}\label{shannon_prob}\quad,
\end{eqnarray}
with the \emph{Shannon entropy} $H=H(X)$ of the letter ensemble $X$ being defined as
\begin{equation}
	H(X):=-\sum_{x}p(x)\log p(x)\quad.
\end{equation}
The probability of a message being in the typical set is given by
\begin{equation}
	P_T:=P(x^N\in T_\delta^N)=\sum_{x^N\in T_\delta^N}p(x^N)\quad,
\end{equation}
and the number of typical messages obeys
\begin{equation}\label{shannon_typ}
	(1-\epsilon)2^{N(H-\delta)}\leq\ |T_\delta^N|\ \leq2^{N(H+\delta)}\quad.
\end{equation}
Shannon's noiseless coding theorem states that for all $\epsilon,\delta>0$ there is an $N_0\in\mathbbm N$, such that for any $N>N_0$ we have
\begin{equation}\label{prob_succ}
	P_T>1-\epsilon\quad.
\end{equation}
So if we only encode the typical messages and forget the rest, the probability of success, which is given by~(\ref{prob_succ}), will still be satisfying for $N$ being large enough. In that case, the typical set contains approximately
$|T_\delta^N|\approx 2^{NH(X)}$
messages to be encoded. The block compression code maps every typical message to a binary string. Since there are $2^{NH(X)}$ distinct messages to be encoded, one needs for every message $\boldsymbol x$ a binary string of length
$L_c(x^N)\approx NH(X)$.
Untypical messages can be all mapped to the same arbitrary ``junk string''. In real life the length of the codewords is given by the integer next above $NH(X)$. Nevertheless, let us view $L_c(x^N)$ as an ``ideal length'' and accept it to be not an integer. Then the ensemble information of the encoded messages~(\ref{code_ensemble_inf}) equals the information content of each encoded message:
\begin{equation}
	\overline I_c(X^N)=\sum_{x^N}p(x^N)
	L_c(x^N)\log 2=NH(X)\quad.
\end{equation}
The ensemble information~(\ref{ensemble_inf}) of the source messages reads
\begin{equation}
	\overline I(X^N)=\sum_{x^N}p(x^N)
	L(x^N)\log |{\cal A}|=N\log|{\cal A}|\quad.
\end{equation}
Since $H(X)\leq\log|{\cal A}|$ we have
\begin{equation}
	\overline I_c(X^N)\leq\overline I(X^N)\quad,
\end{equation}
thus condition~(\ref{compression}) is satisfied, the block compression code is indeed compressive.

In other words, Shannon's noiseless coding theorem states that for canonical messages $X^N$ the \emph{information per letter} can be compressed to $H(X)$ bits approximately. It is not possible to compress the messages to fewer than $NH(X)$ bits without increasing the probability of error exponentially with $N$. 
This gives reason to think of the Shannon entropy as some kind of ``core information'' where all redundancy due to statistical predictability has been removed. But note that block compression only applies to \emph{canonical messages}. In case of general messages it makes no sense to speak of information \emph{per letter}, since Alice can chose entire strings of arbitrary length with some \emph{a priori} probability, so the notion of a letter ensemble $X$ becomes meaningless. Furthermore, the block compression code is a \emph{block code} which assigns binary integers to entire blocks of strings, which requires a high computational effort. Keep also in mind that it is a \emph{lossy} code where  information may be irreversibly lost, so it is really not a good idea to compress a hard disk by block compression.

\subsection{Variable length compression (Huffman coding)}\label{huffman}

Shannon demonstrated in \cite{Shannon48} that a message ensemble may be losslessly compressed by adapting the codeword lengths to the probability of the messages. The average length per symbol of the encoded messages will, in the optimal case, approach the Shannon entropy. Instead of encoding entire messages at once, one can design a symbol code that encodes each letter separately such that optimal compression is achieved by variable codeword lengths. Of such kind is the \emph{Huffman code}, which is a binary symbol coding scheme that applies to messages of arbitrary length and is optimal on canonical messages. Furthermore, it is a \emph{lossless prefix code}, so any source message can be retrieved from its encoding instantaneously and without any loss of information. The Huffman code is completely defined by a single-letter code $c:{\cal A}\rightarrow {\cal A}_C$, mapping each letter $x$ to a binary codeword $c(x)\in{\cal A}_C\subset\{0,1\}^+$ of variable length $L_c(x)$. The extended code on strings $x^n$ of arbitrary length $n$ is given by
\begin{equation}
	c(x^n):=c(x_1)\cdots c(x_n)\quad.
\end{equation}
Because the raw information content of each encoded letter $x$ equals the average length, i.e. $I_c(x)=L_c(x)$, the letter ensemble information content is given by 
\begin{equation}
	\overline I_c(X)=\sum_{x}p(x)\,L_c(x)\quad.
\end{equation}
It can be shown by using the Kraft inequality (see \cite{MacKay}, pp 93) that the average length of any lossless binary symbol code fulfills
\begin{equation}
	\overline L_c(X)\geq H(X)
\end{equation}
with equality if and only if 
\begin{equation}
	L_c(x)=-\log p(x)\quad.
\end{equation}
Of course, in real life $L_c$ must be the integer next above $(-\log p(x))$. This is what the Huffman code does. It constructs to every source letter $x$ a binary prefix codeword with a length between $(-\log p(x))$ and $(-\log p(x)+1)$. This way, it is an optimal symbol code on the alphabet ${\cal A}$, minimizing the ensemble informtation of each letter. Again, let us view $(-\log p(x))$ as an ``ideal length'' that can be interpreted as the \emph{core information} of a the letter $x$, where all redundancy due to statistical predicatbility has been removed. It is given the name \emph{Shannon information content}, denoted by
\begin{equation}
	h(x):=-\log p(x)\quad.
\end{equation}
The ensemble average of $h(x)$ yields the \emph{Shannon entropy}
\begin{equation}
	H(X)=\,<h(X)>\,=-\sum_{x}p(x)\log p(x)\quad.
\end{equation}
Now consider messages $x^N$ drawn from the canonic ensemble $X^N$. The information content of each encoded message is
\begin{equation}
	I_c(x^N)=L_c(x^N)=\sum_{n=1}^N L_c(x_n)\quad.
\end{equation}
The ensemble information thus reads
\begin{eqnarray}
	\overline I_c(X^N)&=& N\overline L_c(X)\quad.
\end{eqnarray}
An ideal Huffman code providing $L_c(x)=h(x)$ would give the ensemble information
\begin{equation}
	\overline I_c(X^N)=NH(X)\quad.
\end{equation}
Since $H(X)\leq\log|{\cal A}|$ the Huffman code is a compression code satisfying condition~(\ref{compression}). For any lossless code we have $\overline L_c(X)\geq H(X)$, so it is an optimal lossless code on canonical messages of any length. How about disdvantages? There is some probability that a particular message is \emph{lengthended} instead of being compressed. This is the price for having a lossless code. While a \emph{lossy} code compresses the most probable files but \emph{forgets} the rest, a \emph{lossless} code compresses the most probable files and \emph{enlarges} the rest. In both cases holds: The bigger the message, the less likely the bad case.

\subsection{General variable length compression}\label{gen_huffman}

The principle of variable length coding can be used to compress general messages $\boldsymbol X$. Given a set $\Omega\subset{\cal A}^+$ of messages $\boldsymbol x$ of fixed or variable length over the alphabet ${\cal A}$, distributed by $p(\boldsymbol x)$.
Now take the message set $\Omega$ itself for an alphabet, i.e. construct a Huffman code that maps any message $\boldsymbol x\in\Omega$ to a binary codeword of length 
\begin{equation}
	L_c(\boldsymbol x)=-\log p(\boldsymbol x)\quad.
\end{equation}
Then again the ensemble information is minimized to
\begin{equation}
	\overline I_c(\boldsymbol X)=-\sum_{\boldsymbol x\in\Omega} p(\boldsymbol x)
	\log p(\boldsymbol x)=H(\boldsymbol X)\quad.
\end{equation}
Of course, if the message set $\Omega$ is infinite, it would take forever to construct the corresponding Huffman code. But if $\Omega$ is small enough or if the probability distribution is sharply peaked around a small subset of $\Omega$ it might even be more effective to construct a Huffman code on the message set $\Omega$ than on the alphabet ${\cal A}$. However, we lose the advantage of sequentially coding, i.e. coding letter by letter, since the code assigns a codeword to entire messages rather than to each letter.
In case of canonical messages $\boldsymbol X=X^N$ we have
\begin{equation}
	\overline I_c(X^N)=H(X^N)=NH(X)\quad,
\end{equation}
hence by the ``Huffman'' message code the same compression is achieved as by the Huffman symbol code.

\subsection{Core information content}\label{core_inf_cont}

Just like the raw information content of a given message measures the \emph{real} effort of communicating it, the Shannon information content measures the \emph{ideal} effort, after encoding it by an optimal compression code that exploits the statistical properies of the whole ensemble. We may therefore define an observable \emph{core information content} $I_0:\Omega\subset{\cal A}^+\rightarrow [0,\infty)$, applying to general messages $\boldsymbol x$ of an ensemble $\boldsymbol X$, giving each message its Shannon information content:
\begin{equation}
	I_0(\boldsymbol x):= h(\boldsymbol x)=-\log p(\boldsymbol x)\quad.
\end{equation}
The average of $I_0$, the \emph{ensemble core information content}, is equal to the Shannon entropy:
\begin{equation}
	\overline I_0(\boldsymbol X):=H(\boldsymbol X)
	=-\sum_{\boldsymbol x\in\Omega}p(\boldsymbol x)\log p(\boldsymbol x)\quad.
\end{equation} 
Why these new names, since there is nothing new defined? The motivation is to stress out the \emph{meaning} contained in the notions of Shannon information content and Shannon entropy, in order to make a generalization to quantum information possible. It will then appear more reasonable to speak of an observable ``core information content''.

In order to illustrate the difference between the raw information content and the core information content imagine two books. Surely, one has to pay twice the price to buy them, since the printer has twice the work by printing them. If this book can be downloaded from the internet and you would download it twice, it would take twice the time and occupy twice the space on your hard disk. So there is a double \emph{raw} information content of these two books. Though at the very moment you notice your download mistake, you surely would delete one of the copies from your hard disk, since it does not contain twice the \emph{core} information. The two books can be compressed to one book without any loss of information. This is possible by reversibly mapping the message set $\Omega$ of the single book to the set $\Omega_2:=\{(\boldsymbol x,\boldsymbol x)\mid\boldsymbol x\in\Omega\}$ of pairs of messages representing two copies of the book, and \emph{vice versa}. As this is a lossless code and the probability distributions on $\Omega$ and $\Omega_2$ are identical, the core information of one book equals the core information of two copies of the same book.

\section{Quantum information theory}

\subsection{Notions and definitions}\label{notions}

For further details on the following notions and definitions the reader is  referred to~\cite{Bostroem_mlconcepts}.  A \emph{quantum alphabet} ${\cal Q}$ is a set of Hilbert vectors normalized to unity,
\begin{equation}
	{\cal Q}:=\{|x\rangle\}\subset{\cal H}\quad.
\end{equation}
The \emph{letters} of ${\cal Q}$ span the \emph{letter space} ${\cal H}_{\cal Q}:=\text{Span}({\cal Q})$.
Since the letter states do neither have to be mutually orthogonal nor linearly independent, the dimension of the letter space reads in general
$K_{\cal Q}:=\dim{\cal H}_{\cal Q}\leq|{\cal Q}|$, 
with equality if the letter states are linearly independent. There is a set of mutually orthogonal \emph{basis letters} ${\cal B}_{\cal Q}=\{|a\rangle\}_a$ with $\dim{\cal H}_{\cal Q}=|{\cal B}_{\cal Q}|$.
A \emph{quantum string} is a product vector $|x^n\rangle=|x_1\rangle\otimes\ldots\otimes|x_n\rangle$ of letter states $|x\rangle$. All possible quantum strings over the alphabet ${\cal Q}$ form the set
\begin{equation}
	{\cal Q}^n:=\{|x^n\rangle\}\subset{\cal H}^n\quad,
\end{equation}
The elements of ${\cal Q}^n$ span the \emph{block space}
\begin{equation}
	{\cal H}_{\cal Q}^n:=\text{Span}({\cal Q}^n)\quad.
\end{equation}
where ${\cal H}_{\cal Q}^0:=\text{Span}({\cal Q}^0)$ is the one-dimensional space spanned by the empty message $|\cdot\rangle$ that forms the set ${\cal Q}^0:=\{|\cdot\rangle\}$.
A \emph{many-letter message} is a vector $|\varphi\rangle$ in the \emph{many-letter space}
\begin{eqnarray}
	{\cal M}_{\cal Q}&:=&\bigoplus_{n=0}^\infty{\cal H}_{\cal Q}^n\quad,
\end{eqnarray}
and can generally be represented as a superposition of block strings $|a^n\rangle$ over the basis alphabet ${\cal B}_{\cal Q}$:
\begin{equation}
	|\varphi\rangle=\sum_{n=0}^\infty\sum_{a^n}\varphi(a^n)\,|a^n\rangle\quad,
\end{equation}
with the wave components $\varphi(a^n):=\langle a^n|\varphi\rangle$ having distinct length $n$. 
An \emph{a priori} message ensemble is represented by a random variable $|\Phi\rangle$, whose realizations are quantum messages $|\varphi\rangle$ chosen from a \emph{source message set} $\Gamma$ with \emph{a priori} probabilities $p(\varphi)$. The corresponding \emph{message matrix} reads
\begin{equation}
	\sigma=\sum_{\varphi\in\Gamma} p(\varphi)\,|\varphi\rangle\langle\varphi|.
\end{equation}	
A \emph{canonical message} is a product message $|x^n\rangle$ chosen from an ensemble $|X^n\rangle$ with probability $p(x^n)=p(x_1)\cdots p(x_n)$.
For canonical messages there is a \emph{letter matrix}
\begin{equation}
	\rho=\sum_x p(x)\,|x\rangle\langle x|,
\end{equation}
such that the message matrix separates into the $n$-fold tensor product of the letter matrix, i.e. $\sigma=\rho^{\otimes n}$. A \emph{grand canonical message} is represented by the message matrix
\begin{equation}
	\sigma=\sum_{n=0}^\infty\lambda_n\,\rho^{\otimes n},
\end{equation}
with $\lambda_n\geq0$, $\sum_n\lambda_n=1$.
The length of a message can be observed by
the self-adjoint \emph{length operator} $\widehat L$ acting on the many-letter space ${\cal M}_{\cal Q}$, represented by a spectral decomposition of mutually orthogonal projectors $\Pi_n$ on ${\cal M}_{\cal Q}$, such that
\begin{equation}
	\widehat L=\sum_{n=0}^\infty n\, \Pi_n \quad,
\end{equation}
with
\begin{equation}
	\Pi_n\,\Pi_m=\delta_{nm}\Pi_n,\quad
	\sum_{n=1}^\infty  \Pi_n ={\mathbbm1}\quad.
\end{equation}
The eigenspaces of the length operator are the block message spaces ${\cal H}_{\cal Q}^n$, which are subspaces of the many-letter space ${\cal M}_{\cal Q}$. Hence the eigenvalues $n$ of $\widehat L$ are degenerate by
$K_{\cal Q}^n:=\dim{\cal H}_{\cal Q}^n=(\dim{\cal H}_{\cal Q})^n$. Using the basis letter set ${\cal B}_{\cal Q}=\{|a\rangle\}_a$ one obtains the spectral decomposition
\begin{equation}
	{\mathbbm 1}=\sum_{n=0}^\infty\sum_{a^n}|a^n\rangle\langle a^n|
\end{equation}
of the unity operator on ${\cal M}_{\cal Q}$. The sum above is always understood as the sum over all distinct strings of length $n$ over the basis alphabet ${\cal B}_{\cal Q}=\{|a\rangle\}$.

\subsection{Raw quantum information content}

It is tempting (and we will give in to this temptation) to define an observable that measures the \emph{quantum information} contained in a single message $|\varphi\rangle$. Bob builds a meter on top of his receiver that he can switch on and which then indicates the number of active quantum subsystems while receiving a message from Alice. Say, Alice sends him a single-letter message $|x\rangle$ from the quantum alphabet ${\cal Q}$. In order to receive this message, the receiver has to be sensible enough to recognize each wave component of the message, whose number is $\dim{\cal H}_{\cal Q}$. If Alice sends him a block message of length $n$, there are $n$ receiver units in action. Bob calibrates his meter to show 1 unit of quantum information if Alice sends him a message composed from a two-state system. 
In analogy to the reasoning of section~\ref{inf_cont}, the quantum information content of a block message $|\varphi\rangle$ of length $L(\varphi)$ composed from the alphabet ${\cal Q}$ reads
\begin{equation}
	I(\varphi)=\log(\dim{\cal H}_{\cal Q})\,L(\varphi)\quad.
\end{equation} 
Consequently, the observable that measures the \emph{raw quantum information content} of an arbitrary quantum message $|\varphi\rangle\in{\cal M}_{\cal Q}$ can be defined as
\begin{equation}
	\widehat I:=\log(\dim{\cal H}_{\cal Q})\,\widehat L\quad.
\end{equation}
Using the orthogonal letter basis ${\cal B}_{\cal Q}=\{|a\rangle\}$, the length operator may be written as
\begin{equation}
	\widehat I=\log(\dim{\cal H}_{\cal Q})
	\sum_{n=0}^\infty\sum_{a^n} n\,
	|a^n\rangle\langle a^n|\quad.
\end{equation}
It is typically quantum that a given message has generally no well-defined information content. Rather, there is an \emph{expected raw quantum information content}, given by
\begin{equation}
	I(\varphi)=\langle\varphi|\widehat I|\varphi\rangle\quad.
\end{equation}
Like every measurement, the detection of its quantum information content potentially disturbs the message. The number ``quantum information content'' is itself a classical information that destroys quantum correlations between components of distinct information content.
 
The \emph{(expected) raw quantum information content} of an arbitrary message matrix $\sigma$ is calculated by
\begin{equation}\label{qinf_cont_ens}
	 I(\sigma):=\text{Tr}\{\sigma\widehat I\}\quad.
\end{equation}
The unit of quantum information is ``1 qbit''.
So, within the presented framework, the name ``qbit'' obtains two meanings: 1) A two-level quantum system and 2) The unit of quantum information, measured by the observable $\widehat I$. This goes in close analogy to classical information theory, where the name "bit" also means both a two-state system (e.g. a dot on a compact disk) and the unit of classical information.

\section{Quantum Coding}

\subsection{Encoding}

A classical code is a function that maps one message ensemble to another. Thus a quantum code simply maps one quantum message ensemble to another. Let the \emph{source ensemble} be an ensemble of general many-letter messages, defined by the random variable
\begin{equation}
	|\Phi\rangle:=\{[|\varphi\rangle,p(\varphi)]\mid|\varphi\in\Gamma\}\quad,
\end{equation} 
with the \emph{source set}
\begin{equation}
	\Gamma:=\{|\varphi\rangle\in{\cal M}_{\cal Q}\mid p(\varphi)>0\}
\end{equation}
of many-letter messages composed from the alphabet ${\cal Q}=\{|x\rangle\}$.
The source message ensemble corresponds to the \emph{source matrix}
\begin{equation}
	\sigma:=\sum_{\varphi\in\Gamma}p(\varphi)|\varphi\rangle\langle\varphi|\quad.
\end{equation}
A code maps the source ensemble to a \emph{code ensemble} $|\Phi_C\rangle$ of code messages $|\varphi_C\rangle$ over a code alphabet ${\cal Q}_C$, taken from a code set $\Gamma_C$ with \emph{a priori} probabilities $p_C(\varphi_C)$. The code alphabet spans another code letter space ${\cal H}_C$
which in turn induces a many-letter code space ${\cal M}_C$ containing all messages that can be composed from ${\cal Q}_C$.
The code ensemble is also represented by the \emph{code matrix} $\sigma_C\in{\cal S}({\cal M}_C)$ which is required to be a density matrix. 
The code $c$ can be represented by a superoperator $\check c$ acting on the state space ${\cal S}({\cal M}_{\cal Q})$ of density matrices over the many-letter space ${\cal M}_{\cal Q}$ and mapping them into the state space ${\cal S}({\cal M}_C)$ of encoded density matrices over the many-letter space ${\cal M}_C$. Thus we have $\check c:{\cal S}({\cal M}_{\cal Q})\rightarrow{\cal S}({\cal M}_C)$ with $\sigma_C=\check c(\sigma)$.

The most general thing that Alice can do do with the quantum state $\sigma$ is a \emph{completely positive map (CPM)}, i.e. a completely positive function mapping density matrices to density matrices. Every CPM has a \emph{Kraus representation}
\begin{equation}\label{kraus_rep}
	\sigma_C=\sum_i E_i\,\sigma\,E_i^\dagger
\end{equation}
with \emph{Kraus operators} $E_i$. The CPM needed here maps states in ${\cal S}({\cal M}_{\cal Q})$ to states in a different state space ${\cal S}({\cal M}_C)$. So the Kraus operators governing the encoding process are linear operators $E_i:{\cal M}_{\cal Q}\rightarrow{\cal M}_C$, which are named \emph{encoders}, fulfilling the \emph{Kraus property}
\begin{equation}\label{kraus_cond}
		\sum_i E_i^\dagger E_i={\mathbbm 1}
\end{equation}
A code having the additional property
\begin{equation}\label{unital}
	\sum_i E_i E_i^\dagger={\mathbbm 1}_{{\cal M}_C}
\end{equation}
is a \emph{unital} code.

Alice encodes each of her \emph{a priori} messages $|\varphi\rangle$ into a generally mixed state
\begin{equation}
	\sigma_\varphi^C:=\sum_i E_i\,|\varphi\rangle\langle\varphi|E_i^\dagger
	\quad,
\end{equation}
so Bob will receive the encoded message ensemble
\begin{equation}\label{mixed_phi}
	\sigma_C=\sum_{\varphi\in\Gamma}p(\varphi)\,\sigma_\varphi^C
	=\sum_i E_i\,\sigma\,E_i^\dagger\quad.
\end{equation}

\subsection{Decoding}

Bob wants to decode the encoded message he obtained from Alice. 
He applies a CPM, given by some Kraus operators $D_j:{\cal M}_C\rightarrow{\cal M}_{\cal Q}$, called \emph{decoders}, and finally obtains the \emph{decoded matrix}
\begin{equation}\label{decoded}
	\begin{split}
		\sigma'&:=\sum_{ij}D_j E_i\,\sigma\,E_i^\dagger D_j^\dagger
		=\sum_{\varphi\in\Gamma}p(\varphi)\,\sigma_\varphi\quad,
	\end{split}
\end{equation}
where $\sigma_\varphi$
is the mixed state that Bob obtains by decoding the encoded \emph{a priori} state $|\varphi\rangle$.
In general the decoded matrix $\sigma'$ is not identical to the source matrix $\sigma$. What can be said about the confidence of the transmission? Say, Alice sends a message $|\varphi\rangle$. After encoding and decoding, the message will be crumbled into the mixed state 
\begin{equation}
	\sigma_\varphi=\sum_{ij}D_j E_i\,|\varphi\rangle\langle\varphi|
	\,E_i^\dagger D_j^\dagger\quad.
\end{equation}
Though still there is a certain probability for Bob that he can recover the original message by a generalized measurement. The probability of finding the state $|\varphi\rangle$ in the ensemble $\sigma'$ is given by the \emph{fidelity}
\begin{equation}
	F(\varphi)=\langle\varphi|\sigma_\varphi|\varphi\rangle
	=\sum_{ij}\big|\langle\varphi|D_j E_i|\varphi\rangle\big|^2.
\end{equation}
The \emph{confidence} of the code is then defined by the average fidelity,
\begin{equation}\label{confidence}
	\overline F:=\sum_{\varphi\in\Gamma}p(\varphi)\sum_{ij}
	|\langle\varphi|D_j E_i|\varphi\rangle|^2\quad.
\end{equation}
whereas the \emph{probability of error} is given by
\begin{equation}\label{prob_error}
	P_{err}:=1-\overline F\quad.
\end{equation}
Bob now has the task to construct decoders $D_j$ optimizing the confidence of the code, i.e. decreasing the probability of error.
However, the confidence cannot be expressed in terms of density matrices. It is an expression that requires Alice's \emph{a priori} knowledge of the message ensemble, i.e. the random variable $|\Phi\rangle$. So Bob has to do the job \emph{together} with Alice, constructing suitable decoders that maximize the confidence of the transmission. 

\subsection{Lossy and lossless codes}

A \emph{lossless} code is represented by an invertible superoperator $\check c$ with only one Kraus operator $E$. According to~(\ref{kraus_cond}), $E$ must be an \emph{isometric} operator, i.e. $E^\dagger E={\mathbbm1}$.
A \emph{unitary} code fulfills in addition
$E E^\dagger={\mathbbm1}_{{\cal M}_C}$.
Using a lossless code, Alice encodes her source matrix through
$\sigma_C=E\,\sigma\,E^\dagger$,
and Bob decodes it uniquely through
$\sigma=E^\dagger\,\sigma_C\,E$.
For a lossless code the confidence of transmission, given by~(\ref{confidence}), is $\overline F=1$.

A \emph{lossy} code has a Kraus representation with more than one Kraus operator. It is not possible to uniquely recover the source matrix $\sigma$. Instead, the decoding process, using decoders $D_j$, gives a \emph{decoded matrix} $\sigma'$ given by~(\ref{decoded}). 
For a lossy code the confidence of transmission is 
$\overline F<1$.
If the confidence can be made close to unity, the lossy code may be useful.

\subsection{Compression codes}

A quantum compression code reduces the information content of the message ensemble $\sigma$, given by~(\ref{qinf_cont_ens}). The quantum information content of an encoded state is represented by the observable
$\widehat I_C=\log(\dim{\cal H}_C)\,\widehat L_C$,
where $\widehat L_C$ is the length operator in the code space ${\cal M}_C$.
Though it is more convenient to express everything in the source space ${\cal M}_{\cal Q}$.

The average length of an encoded state $\sigma_C\in{\cal S}({\cal M}_C)$ is given by
$L_C(\sigma_C)=\text{Tr}\{\sigma_C\widehat L_C\}$,
where the encoded state is obtained from the original state $\sigma\in{\cal S}({\cal M}_{\cal Q})$ by
$\sigma_C=\sum_i E_i\,\sigma\,E_i^\dagger$.
The length operator $\widehat L_C$ on ${\cal M}_C$ can be mapped to an observable $\widehat L_c$ on ${\cal M}_{\cal Q}$ by
\begin{equation}
	\widehat L_c:=\sum_i E_i^\dagger\,\widehat L_C\,E_i\quad,
\end{equation}
such that the average of $\widehat L_c$ for the source ensemble $\sigma$,
$L_c(\sigma)=\text{Tr}\{\sigma\,\widehat L_c\}$,
equals the ensemble length $ L_C(\sigma_C)$ of the encoded ensemble $\sigma_C$.
That way, one can define the observable \emph{encoded quantum information}, acting on the source space ${\cal M}_{\cal Q}$, by
\begin{equation}\label{enc_inf}
	\widehat I_c:=\log(\dim{\cal H}_C)\,\widehat L_c\quad,
\end{equation}
so that the expected encoded quantum information content of a message matrix $\sigma$ reads 
\begin{equation}
	I_c(\sigma)=\text{Tr}\{\sigma\widehat I_c\}\quad.
\end{equation}
The observable $\widehat I_c$ indicates how long a source message would be if it was encoded by $c$. A \emph{compression code} is thus a code $\check c:{\cal S}({\cal M}_{\cal Q})\rightarrow{\cal S}({\cal M}_C)$ that reduces the quantum information of the message ensemble, i.e.
\begin{equation}\label{comp_qcode}
	 I_c(\sigma)\leq  I(\sigma)\quad.
\end{equation}

\subsection{Translation of messages}\label{translation}

Alice has just typed a message to Bob into her quantum computer and now wants to save it. But the quantum hard disk only operates with qbits, whereas the message is written in english. So the quantum computer has to invoke an algorithm to \emph{translate} the message from the english alphabet to the qbit alphabet. Needless to say, lossless coding is desired here. To put it more general,
let ${\cal Q}, {\cal Q}_C$ be two quantum alphabets with corresponding basis alphabets ${\cal B}_{\cal Q}=\{|a\rangle\}_a$, ${\cal B}_C=\{|c\rangle\}_c$, spanning the letter spaces ${\cal H}_{\cal Q}$, ${\cal H}_C$ and inducing the many-letter spaces ${\cal M}_{\cal Q}$, ${\cal M}_C$, respectively.
A \emph{translation code} between the alphabets ${\cal Q}$ and ${\cal Q}_C$
is completely specified by an isometric \emph{block translator} $\hat t:{\cal H}_{\cal Q}^N\rightarrow{\cal H}_C^M$ mapping each block of $N$ source basis letters to a block of $M$ code basis letters, i.e.
\begin{equation}
	\forall |a^N\rangle\in{\cal B}_{\cal Q}^N:\quad
	\hat t\,|a^N\rangle:=|c^M(a^N)\rangle\in{\cal B}_C^M\quad,
\end{equation}
where
$|c^M(a^N)\rangle=|(c_1\cdots c_M)(a^N)\rangle$
is a string of $M$ basis letters over the code alphabet with
$\langle c^M(a^N)|c^M(a{'}^N)\rangle=\delta_{a^Ma{'}^M}$.
The value
\begin{equation}
	R:=\frac{M}{N}
\end{equation}
is called the \emph{rate} of the code and has to fulfill
\begin{equation}\label{rate_cond}
	R\geq \frac{\log(\dim{\cal H}_{\cal Q})}{\log(\dim{\cal H}_C)}
\end{equation}
in order to reversibly encode each source letter block.

Since the basis letters are mutually orthogonal, the letter translator $\hat t$ reads
\begin{equation}\label{letter_trans}
	\hat t=\sum_{a^N}
	| c^{NR}(a^N)\rangle\langle a^N|\quad.
\end{equation}
The \emph{message translator} is then defined by
\begin{equation}\label{message_trans1}
	\widehat T:=\sum_{n=0}^\infty \hat t^{\otimes n}\quad,
\end{equation}
where
\begin{eqnarray}
	\hat t^{\otimes 0}&:=&|\cdot\rangle\langle\cdot|\\
	\hat t^{\otimes n}&=&\hat t\otimes\cdots\otimes\hat t\quad.
\end{eqnarray}
Because the block translator $\hat t$ is isometric, the message translator $T:{\cal M}_{\cal Q}\rightarrow{\cal M}_C$ is also isometric, i.e.	$T^\dagger T={\mathbbm1}$, and reads in general
\begin{equation}\label{message_trans}
	T=\sum_{n=0}^\infty\sum_{a^{nN}} |c^{nNR}(a^{nN})\rangle\langle a^{nN}|\quad,
\end{equation}
where $|a^{nN}\rangle=|a_1^N\cdots a_n^N\rangle$ denotes a string of $n$ blocks of length $N$ being mapped to a codeword 
$|c^{nNR}\rangle=|c_1^{NR}\cdots c_n^{NR}\rangle$ of $n$ blocks of length $NR$.
Every quantum message $|\varphi\rangle\in{\cal M}_{\cal Q}$ is translated to 
\begin{equation}
	T|\varphi\rangle=\sum_{n=0}^\infty\sum_{a^{nN}}
	\varphi(a^{nN})\,|c^{nNR}(a^{nN})\rangle\quad,
\end{equation}
with the wave components
$\varphi(a^{nN}):=\langle a^{nN}|\varphi\rangle$.
The whole message ensemble $\sigma\in{\cal S}({\cal M}_{\cal Q})$ is translated to $\sigma_C=T\,\sigma\,T^\dagger$.

The observable measuring the encoded quantum information of a message being translated is according to~(\ref{enc_inf})
\begin{eqnarray}
	\widehat I_c&=&\log(\dim{\cal H}_C)\widehat L_c
	=\log(\dim{\cal H}_C)\widehat T^\dagger\,\widehat L_C\,\widehat T\\
	&=&\log(\dim{\cal H}_C)\sum_{n=0}^\infty\sum_{a^{nN}} nNR\,
	|a^{nN}\rangle\langle a^{nN}|\\
	&=&R\,\frac{\log(\dim{\cal H}_C)}{\log(\dim{\cal H}_{\cal Q})}
	\widehat I\quad.
\end{eqnarray}  
Since the rate $R$ has to fulfill condition~(\ref{rate_cond}), we have
$\widehat I_c\geq \widehat I$,
i.e. translation codes are never compressive. This is reasonable since compression is only possible by taking advantage of statistical properties of the message ensemble. A translation code is not based on statistical properties, hence it cannot be compressive. In the best case, the rate fulfills~(\ref{rate_cond}) with equality, so the encoded raw information just equals the source information.\medskip

\textbf{Case 1: }$\dim{\cal H}_{\cal Q}\leq \dim{\cal H}_C$.\\
Alice's alphabet is not bigger than the alphabet of the quantum hard disk. So she can chose a block of size $N$ of source letters that is mapped to a single code letter. The rate of the code is $R\leq1$.

\textbf{Case 2: }$\dim{\cal H}_{\cal Q} > \dim{\cal H}_C$.\\
Alice's alphabet is bigger than the alphabet of the quantum hard disk. So it is necessary to find codewords of length $R>1$ for every source basis letter.

\subsection{Block compression: Schumacher coding}\label{schumacher}

\subsubsection{Standard Schumacher coding}

The Schumacher code (see \cite{Schumacher96}) is the quantum analogue to block compression (see section~\ref{block_comp}). It is a lossy code on canonical messages of fixed length $N$. Thus throughout this section we stay in the block space ${\cal H}_{\cal Q}^N$.

Alice uses a canonical message ensemble given by
\begin{equation}
	|X^N\rangle=\{[|x^N\rangle,p(x^N)]\mid |x^N\rangle\in\Gamma\}
\end{equation}
with the source set 
$\Gamma$ of all quantum strings $|x^N\rangle$ of length $N$ over the alphabet ${\cal Q}$ which spans the letter space ${\cal H}_{\cal Q}$. The \emph{a priori} probabilities read
$p(x^N)=p(x_1)\cdots p(x_N)$.
The source message ensemble corresponds to the message matrix
\begin{equation}
	\sigma=\rho^{\otimes N}\equiv \rho\otimes\cdots\otimes \rho\quad,
\end{equation}
where the letter matrix is given by
\begin{equation}
	\rho=\sum_x p(x)|x\rangle\langle x|\quad.
\end{equation}
The set of $\rho$-eigenvectors form a basis letter set ${\cal B}_{\cal Q}=\{|a\rangle\}_a$, such that $\dim{\cal H}_{\cal Q}=|{\cal B}_{\cal Q}|$ and
\begin{equation}
	\rho=\sum_a q(a)|a\rangle\langle a|\quad.
\end{equation}
Hence the source message matrix obtains a diagonal form in the basis strings $|a^n\rangle\in{\cal B}_{\cal Q}^n$ with $q(a^n)=q(a_1)\cdots q(a_n)$.
The ensemble that Alice submits appears to Bob as a mixture of strings over an alphabet ${\cal B}_{\cal Q}$ of perfectly distinguishable letters $|a\rangle$, each one distributed independently by $q(a)$. Shannon's noiseless coding theorem may be applied as follows. 
There is a typical set $T_\delta^N$ of quantum strings $|a^N\rangle$, whose probabilities fulfill~(cf~(\ref{shannon_prob}))
\begin{equation}
	2^{N(H+\delta)}<q(a^N)<2^{-N(H+\delta)}\quad,
\end{equation}
such that for every $\epsilon,\delta>0$ we have
\begin{equation}\label{shannon_succ}
	P_T:=P(|a^N\rangle\in T_\delta^N)> 1-\epsilon\quad.
\end{equation}
Here $H$ is the Shannon entropy of the basis letter ensemble,
\begin{equation}
	H:= -\sum_a p(a)\log p(a)\quad,
\end{equation}
which equals the von Neumann entropy of the letter matrix $\rho$,
\begin{equation}
	S(\rho):=-\text{Tr}\{\rho\log \rho\}\quad,
\end{equation}
i.e. $H=S(\rho)$.
The von Neumann entropy is bounded from above by
\begin{equation}\label{vN_bound}
	S(\rho)\leq \log(\dim{\cal H}_{\cal Q})\quad.
\end{equation}
As the typical set $T_\delta^N$ contains mutually orthogonal vectors, they span a \emph{typical subset}
\begin{equation}\label{typ_subset}
	V_\delta^N:=\text{Span}(T_\delta^N)
\end{equation}
with $\dim V_\delta^N=|T_\delta^N|$. According to Shannon (cf~(\ref{shannon_typ})) we therefore have
\begin{equation}\label{shannon_dim}
	(1-\epsilon)2^{N(S-\delta)}\leq \dim V_\delta^N\leq2^{N(S+\delta)}\quad.
\end{equation}	
Define the projector on the typical subspace by
\begin{equation}
	\Pi_T:=\sum_{|a^N\rangle\in T_\delta^N}|a^N\rangle\langle a^N|\quad,
\end{equation}
then the total probability of messages lying in the typical subspace reads
\begin{equation}
	P_T=\sum_{|a^N\rangle\in T_\delta^N}p(a^N)
	=\text{Tr}\{\rho^{\otimes N}\Pi_T\}\quad,
\end{equation}
so together with~(\ref{shannon_succ}) we have
\begin{equation}\label{schumacher_fid}
	\text{Tr}\{\rho^{\otimes N}\Pi_T\}>1-\epsilon\quad.
\end{equation}
Alice now encodes message components in the typical subspace by the encoder
\begin{equation}
	E_T:=\sum_{|a^N\rangle\in T_\delta^N}|c^R(a^N)\rangle\langle a^N|\quad,
\end{equation}
where $|c^R(a^N)\rangle$ is a unique codeword of length $R$ over an orthogonal code alphabet ${\cal B}_C$ for the typical message $|a^N\rangle$. Since there are $\dim V_\delta^N$ orthogonal messages to encode, the rate $R$ of the code, which gives the dimension of the code space ${\cal H}_C^R$, obeys
\begin{equation}\label{schumacher_rate}
	R\geq\frac{\log(\dim V_\delta^N)}{\log(\dim{\cal H}_C)}\quad,
\end{equation}
where ${\cal H}_C$ is the letter space spanned by the code alphabet ${\cal B}_C=\{|c\rangle\}$. 
Alice maps the components outside the typical subspace to a junk string $|c_{junk}^R\rangle\in{\cal H}_C^R$ of length $R$ orthogonal to the code image of the typical subspace by the encoder
\begin{equation}
	E_{\neg T}:=\sum_{a^N\notin T_\delta^N}|c_{junk}^R\rangle\langle a^N|\quad,
\end{equation}
which gives the second Kraus operator.
Altogether, any \emph{a priori} source message $|x^N\rangle$ is encoded into
the mixed state
\begin{eqnarray}
	\sigma_{x^N}^C&=&E_T|x^N\rangle\langle x^N|E_T^\dagger
	+E_{\neg T}|x^N\rangle\langle x^N|E_{\neg T}^\dagger\\
		&=&\sum_{|a^N\rangle\in T_\delta^N}|\langle a^N|x^N\rangle|^2\,
		|c^R(a^N)\rangle\langle c^R(a^N)|\nonumber\\
		&&+\sum_{a{'}^N\notin T_\delta^N}|\langle a{'}^N|x^N\rangle|^2\,
		|c_{junk}^R\rangle\langle c_{junk}^R|
\end{eqnarray}
Bob decodes the message by applying the decoders
\begin{equation}
	D_T:=E_T^\dagger,\quad
	D_{\neg T}:=\sum_{|c^R\rangle\notin W_T}|a_{junk}^N\rangle
	\langle c^R|,
\end{equation}
where $W_T\subset{\cal H}_C^R$ is the code image of the typical subspace, i.e.
\begin{equation}
	W_T:=c(V_\delta^N),
\end{equation} 
and $|c^R\rangle$ are mutually orthogonal strings of $R$ code basis letters, and $|a_{junk}^N\rangle$ is a junk string of length $N$ outside the typical subspace.
After encoding and decoding the message $|x^N\rangle$ that Alice originally has sent, will be a mixture
\begin{eqnarray}
	\sigma_{x^N}&=&D_T\,\sigma_{x^N}^C\,D_T^\dagger
	+D_{\neg T}\,\sigma_{x^N}^C\,D_{\neg T}^\dagger\\
	&=&\sum_{|a^N\rangle\in T_\delta^N}|\langle a^N|x^N\rangle|^2\,
		|a^N\rangle\langle a^N|\nonumber\\
		&&+\sum_{a{'}^N\notin T_\delta^N}|\langle a{'}^N|x^N\rangle|^2\,
		|a_{junk}^N\rangle\langle a_{junk}^N|\quad.
\end{eqnarray}
How ablout the confidence? The fidelity of  $|x^N\rangle$ in the mixture $\sigma_{x^N}$ reads
\begin{eqnarray}
	F(x^N)&=&\langle x^N|\sigma_{x^N}|x^N\rangle
	=\sum_{|a^N\rangle\in T_\delta^N}|\langle a^N|x^N\rangle|^4\\
	&=&\|\Pi_T|x^N\rangle\|^4
\end{eqnarray}
Since any real number $x$ satisfies $x^2\geq 2x-1$, we have
\begin{equation}
	F(x^N)\geq 2\langle x^N|\Pi_T|x^N\rangle-1\quad.
\end{equation}
It follows for the confidence $\overline F$ of the code:
\begin{eqnarray}
	\overline F&=&\sum_{x^N}p(x^N)F(x^N)
	=\sum_{x^N}p(x^N)\|\Pi_T|x^N\rangle\|^4\\
	&\geq&2\text{Tr}\{\rho^{\otimes N}\Pi_T\}-1\quad.
\end{eqnarray}
Using~(\ref{schumacher_fid}) we conclude that the confidence of the Schumacher code is bounded from below by
\begin{equation}
	\overline F>1-2\epsilon\quad.
\end{equation}
So Alice can achieve arbitrary good confidence by chosing the block size $N$ large enough.

The observable measuring the content of quantum information in a Schumacher encoded message reads according to~(\ref{enc_inf})
\begin{eqnarray}
	\widehat I_c&=&\log(\dim{\cal H}_C)\widehat L_c\\
	&=&\log(\dim{\cal H}_C)\Big[E_T^\dagger\,\widehat L_C\,E_T
	+E_{\neg T}^\dagger\, L_C\,E_{\neg T}\Big]\quad.
\end{eqnarray}
Since we have
\begin{eqnarray}
	\widehat L_C\,|c^R(a^N)\rangle&=&R\,|c^R(a^N)\rangle\\
	\widehat L_C\,|c_{junk}^R\rangle&=&0\quad,
\end{eqnarray}
the encoded information operator reads
\begin{equation}
	\widehat I_c=R\,\log(\dim{\cal H}_C)\,\Pi_T\quad,
\end{equation}
where the rate $R$ fulfills~(\ref{schumacher_rate}).
Although $R$ must be an integer, we consider an \emph{ideal rate} $R$ fulfilling~(\ref{schumacher_rate}) with equality. Furthermore, for $N$ very large, the dimension on $V_\delta^N$ approaches
$\dim V_\delta^N\approx 2^{NS(\rho)}$.
Hence the encoded information reads approximately
\begin{equation}
	\widehat I_c\approx N\,S(\rho)\,\Pi_T\quad,
\end{equation}
whereas the information content of the source messages is measured by
\begin{eqnarray}
	\widehat I&=&\log(\dim{\cal H}_{\cal Q})\,\widehat L\\
	&=&N\,\log(\dim{\cal H}_{\cal Q}){\mathbbm 1}_N\quad,
\end{eqnarray}
where ${\mathbbm1}_N$ is the unity operator on ${\cal H}_{\cal Q}^N$.
Since the von Neumann entropy fulfills~(\ref{vN_bound}) we have
\begin{equation}
	\widehat I\geq \widehat I_c\quad,
\end{equation}
i.e. the Schumacher code is compressive on the \emph{entire} block space ${\cal H}_{\cal Q}^N$, because it fulfills~(\ref{comp_qcode}) for any source message ensemble $\sigma$. This is not surprising, since lossy codes throw away information, hence any source message ensemble can only either be compressed or keep its size. Canonical messages of length $N$, containing $N\,\log(\dim{\cal H}_{\cal Q})$ qbits of information are optimally compressed to $NS(\rho)$ qbits of information. Thus here the quantum information \emph{per letter} is compressed from $\log(\dim{\cal H}_{\cal Q})$ to $S(\rho)$ qbits.
This is not necessarily valid for messages of other types. In the next section, we will extend the Schumacher code to messages of a more general form, namely to \emph{grand canonical messages}, and obtain a similiar result.

\subsubsection{Generalized Schumacher coding}

Within the framework of many-letter theory the Schumacher coding scheme can be generalized to grand canonical messages, i.e. messages $\sigma$ of the form
\begin{equation}
	\sigma=\sum_{n=0}^\infty \lambda_n\,\rho^{\otimes n}\quad.
\end{equation}
The typical subspaces $V_\delta^n$ are spanned by the typical basis strings $|a^n\rangle$ of length $n$ in the typical set $T_\delta^n$. The rate $r$ of the code components depends on $n$ according to~(\ref{schumacher_rate}) for $R\mapsto r$ and $N\mapsto n$ varying.
The \emph{typical many-letter subspace} $V_\delta$ is given by
\begin{equation}
	V_\delta:=\bigoplus_{n=0}^\infty V_\delta^n\quad,
\end{equation}
with $V_\delta^n$ given by~(\ref{typ_subset}) for $N\mapsto n$ varying.
The encoders that Alice uses, read now
\begin{eqnarray}
	E_T&=&\sum_{n=0}^\infty\sum_{|a^n\rangle\in T_\delta^n}
	|c^r(a^n)\rangle\langle a^n|\\
	E_{\neg T}&=&\sum_{n=0}^\infty\sum_{a^n\notin T_\delta^n}
	|\cdot\rangle\langle a^n|
	\quad,
\end{eqnarray}
where again we set $|a^0\rangle:=|\cdot\rangle$, $\rho^{\otimes 0}:=|\cdot\rangle\langle\cdot|$ and let $T_\delta^0$ contain only the empty message $a^0:=(\cdot)$. The junk message is now allowed to be the empty message $|\cdot\rangle$. Bob's decoders look like
\begin{equation}
	D_T:=E_T^\dagger,\quad 
	D_{\neg T}:=\sum_{m=0}^\infty\sum_{|c^m\rangle\notin W_{T}}
	|\cdot\rangle\langle c^m|,
\end{equation}
where $W_{T}$ is the code image of the typical subspace, i.e. $W_{T}:=c(V_\delta)$ and $|c^m\rangle$ are mutually orthogonal code strings of length $m$.
Since every subspace ${\cal H}_{\cal Q}^n$ of messages of length $n$ is orthogonal to a subspace of messages of different length, encoding and decoding of different subspaces does not interfere. Though Schumacher coding will be only confidental and optimal within the higher dimensional subspaces. Considering ideal rates and a length distribution $\lambda_n$ which support lies mostly in higher dimensional subspaces, the information content will be compressed from
\begin{eqnarray}
	\widehat I&=&\log(\dim{\cal H}_{\cal Q})\,\widehat L\\
	&=&\log(\dim{\cal H}_{\cal Q})\sum_{n=0}^\infty n\,\Pi_n
\end{eqnarray}
to
\begin{eqnarray}
	\widehat I_c&=&R\,\log(\dim{\cal H}_C)\,\Pi_{T^n}\\
	&\stackrel{(\ref{schumacher_rate})}{=}&\log(\dim V_\delta^n)\,\Pi_{T^n}\\
	&\approx&\sum_{n=0}^\infty n\,S(\rho)\,
	\Pi_{T^n}\quad,
\end{eqnarray}
where
\begin{equation}
	\Pi_n=\sum_{a^n}|a^n\rangle\langle a^n|
\end{equation}
is the projector onto the subspace ${\cal H}_{\cal Q}^n$ of length $n$ messages and
\begin{equation}
	\Pi_{T^n}=\sum_{|a^n\rangle\in T_\delta^n}|a^n\rangle\langle a^n|
\end{equation}
is the projector on the typical subspace of length $n$ messages.
Since $S(\rho)\leq\log(\dim{\cal H}_{\cal Q})$ and $\Pi_{T^n}\leq\Pi_n$ we have
\begin{equation}
	\widehat I_c\leq\widehat I\quad,
\end{equation}
i.e. the generalized Schumacher code is a compression code on the entire many-letter space.
For grand canonical messages an optimal compression will be achieved. The raw information content of the source messages then reads
\begin{equation}
	I(\sigma)=\sum_{n=0}^\infty \lambda_n\,n\,
	\log(\dim{\cal H}_{\cal Q})\quad.
\end{equation}
The Schumacher code compresses the raw information content to
\begin{equation}
	I_c(\sigma)\approx\sum_{n=0}^\infty \lambda_n\,n\,S(\rho)\,P_{T^n}
\end{equation}
qbits, where $P_{T^n}=\text{Tr}(\rho^{\otimes n}\Pi_{T^n})$ is the probability of a block message of length $n$ lying in the typical subspace $V_\delta^n$. If the support of the length distribution $\lambda_n$ is on subspaces of dimensions being high enough, the confidence of the code is still acceptable, i.e.
\begin{eqnarray}
	P_T&:=&\text{Tr}\{\sigma\,\Pi_T\}=\sum_{n=0}^\infty 
	\lambda_n P_{T^n}>1-\epsilon
\end{eqnarray}
is achievable for any $\epsilon,\delta>0$. The projector $\Pi_T$ onto the typical many-letter subspace $V_\delta$ is defined by
\begin{equation}
	\Pi_T:=\sum_{n=0}^\infty \Pi_{T^n}\quad.
\end{equation}
Note, however, that for a \emph{given} source message ensemble the fidelity can only be increased by a higher tolerance $\delta$ of the typical subspaces, which results in a bad compression. Only if Alice choses a suitable length distribution, she can achieve both optimal compression and reliable transmission.
In the limit where the support of the length distribution $\lambda_n$ is shifted to $n\rightarrow\infty$ we have $P_T\rightarrow1$ and $\dim V_\delta^n\rightarrow n\,S(\rho)$, hence
\begin{equation}
	I_c(\sigma)\rightarrow\sum_{n=0}^\infty \lambda_n\,n\,S(\rho)\quad.
\end{equation}
In this limit, each of the perfectly distinguishable canonical components $\rho^{\otimes n}$ of $\sigma$ is compressed to $n\,S(\rho)$ qbits. The total compressed message is the sum of the compressed components, weightened by $\lambda_n$.  Hence also for grand canonical messages one can say that the Schumacher code compresses each message to $S(\rho)$ qbits \emph{per letter}. This confirms the result already obtained in the last section. Note, however, that the notion of a compression \emph{per letter} only makes sense in case of (grand) canonical messages. Other types of message cannot be Schumacher compressed, just because for them there is no letter matrix $\rho$. Hence in the context of Schumacher compression the von Neumann entropy has not yet a fundamental meaning. In the next section we will introduce a lossless compression scheme applying to \emph{all} messages, that finally establishes the von Neumann entropy as the amount of core quantum information of any given message ensemble.

\section{Lossless compression}

A compression code always makes use of statistical properties of the source message ensemble. As already stated in section~\ref{comp_codes}, a compression code can be realized in two ways
\begin{itemize}
\item[]\textbf{Type 1 (Lossy): }
	Compress the most probable messages and forget the rest, or
\item[]\textbf{Type 2 (Lossless): }
	Compress the most probable messages and enlarge the rest.
\end{itemize}
Since the latter involve codewords of variable length, they can hardly be realized on block spaces. Nevertheless, an implementation of Huffman coding into quantum information theory based on block spaces has been worked out by Braunstein \emph{et al.} (see \cite{Braunstein98}), but due to the restriction to block spaces this coding scheme it is not a lossless scheme. In the framework of many-letter quantum information theory, however, lossless compression is realizable in the following way.

\subsection{Compressing grand canonical messages}

A symbol quantum code over the alphabet ${\cal Q}$ can be represented by a single-letter encoder
\begin{equation}
	C_{\cal Q}:=\sum_a |c(a)\rangle\langle a|\quad,
\end{equation}
where ${\cal B}_{\cal Q}=\{|a\rangle\}_a$ is a basis letter set spanning the letter space ${\cal H}_{\cal Q}$ and $|c(a)\rangle$ is a string  of code letters taken from an orthogonal code alphabet ${\cal B}_C=\{|c\rangle\}$. 
Thus the length of the codeword is
\begin{equation}
	\widehat L_C\,|c(a)\rangle=L_c(a)\,|c(a)\rangle\quad.
\end{equation}
The extension of the code $c$ to strings of arbitrary length can be given by
\begin{equation}
	|c(a^n)\rangle
	:=|c(a_1)\cdots c(a_n)\rangle\quad,
\end{equation}
so the total length of the encoded message $|a^n\rangle$ reads
\begin{equation}
	\widehat L_C|c(a^n)\rangle
	=L_c(a^n)|c(a^n)\rangle\quad,
\end{equation}
where $L_c(a^n):=L_c(a_1)+\ldots+L_c(a_n)$.
The code must be uniquely decodeable, i.e.
$c(a^n)\neq c(a{'}^m)$ for $a^n\neq a{'}^m$,
so the code messages must fulfill
\begin{equation}
	\langle c(a^n)|c(a{'}^m\rangle=0
	\quad\text{for }a^n\neq a{'}^m\quad.
\end{equation}
The total encoder of all messages is constructed by
\begin{equation}
	C:=\sum_{n=0}^\infty C_{\cal Q}^{\otimes n}\quad,
\end{equation}
where
\begin{eqnarray}
	C_{\cal Q}^{\otimes 0}&:=&|\cdot\rangle\langle\cdot|\\
	C_{\cal Q}^{\otimes n}&:=&C_{\cal Q}\otimes\cdots\otimes C_{\cal Q}\quad,
\end{eqnarray}
i.e. the empty message stays empty and all other strings are encoded letter by letter. The encoder, which can also be written as
\begin{equation}
	C=\sum_{n=0}^\infty\sum_{a^n}|c(a^n)\rangle\langle a^n|\quad,
\end{equation}
is an isometric operator on the many-letter space ${\cal M}_{\cal Q}$, since
\begin{eqnarray}
	C^\dagger C&=&\sum_{n,m=0}^\infty\sum_{a^n,a{'}^m}
	|a^n\rangle\langle c(a^n)|c(a{'}^m)\rangle
	\langle a{'}^m|\\
	&=&\sum_{n=0}^\infty\sum_{a^n}
	|a^n\rangle\langle a^n|={\mathbbm1}\quad.
\end{eqnarray}

Alice now choses her messages from the grand canonical message ensemble \begin{equation}\label{alice_grand}
	\sigma=\sum_{n=0}^\infty\lambda_n\,\rho^{\otimes n}\quad,
\end{equation}
where the letter matrix $\rho$ is given by
\begin{equation}
	\rho=\sum_x p(x)|x\rangle\langle x|\quad,
\end{equation}
with the diagonalization 
\begin{equation}
	\rho=\sum_a q(a)|a\rangle\langle a|\quad,
\end{equation}
i.e. we have chosen the basis alphabet ${\cal B}_{\cal Q}$ such that $\rho$ becomes diagonal. To Bob it appears as if Alice would send him perfectly distinguishable messages $|a^n\rangle$ over the alphabet ${\cal B}_{\cal Q}$ distributed by $q(a^n)=q(a_1)\cdots q(a_n)$. Hence it is a good idea to invoke a Huffman coding scheme (see section~\ref{huffman}) mapping each letter $|a\rangle$ to a binary codeword $|c(a)\rangle$ of length
$L_c(a)=-\log q(a)$.
Again, the above length is in general not an integer and one has to take the integer next above instead. Though we regard the above number as an \emph{ideal length} provided by an ideal Huffman code. Since it is an optimal code, the average length of the encoded letter ensemble is minimized to the Shannon entropy of the basis letter ensemble
\begin{equation}
	\overline L_c=H(A)=-\sum_a q(a)\log q(a)\quad.
\end{equation}
Since the Huffman code is a binary code, the average length equals the average information content of the letter ensemble:
\begin{equation}
	 I_c(\rho)=\log(\dim{\cal H}_C) L(\rho)
	= L_c(\rho)\quad,
\end{equation}
whereas the Shannon entropy of the basis letter ensemble equals the von Neumann entropy of the letter matrix, $H(A)=S(\rho)$, thus we have
$I_c(\rho)=S(\rho)$. Since $S(\rho^{\otimes n})=n\,S(\rho)$, the grand canonical message ensemble~(\ref{alice_grand}) contains 
\begin{equation}
	 I_c(\sigma)=\sum_{n=0}^\infty\lambda_n\,n\,S(\rho)=S(\sigma)
\end{equation}
qbits of encoded information on the average. Since the original information content is
$I(\sigma)=\sum_{n=0}^\infty\lambda_n\,n\,\log(\dim{\cal H}_{\cal Q})$
and since $S(\rho)\leq\log(\dim{\cal H}_{\cal Q})$ this coding scheme is a compressive code according to~(\ref{comp_qcode}).

\subsection{Compressing general messages}

In analogy to section~\ref{gen_huffman} we can introduce a general coding scheme that optimally compresses an arbitrary message ensemble 
\begin{equation}\label{sigma}
	\sigma=\sum_{\varphi\in\Gamma}p(\varphi)\,
	|\varphi\rangle\langle\varphi|
\end{equation}
over a given source alphabet ${\cal Q}=\{|x\rangle\}$ without any loss of information.
Let
\begin{equation}
	\sigma=\sum_i q_i\,|e_i\rangle\langle e_i|
\end{equation}
be a diagonalization of $\sigma$, where the $|e_i\rangle$'s are eigenvectors of $\rho$ to nonzero eigenvalues $q_i>0$ and generally no product messages but rather superpositions of strings $|a^n\rangle$ over some orthogonal basis alphabet ${\cal B}_{\cal Q}=\{|a\rangle\}_a$:
\begin{equation}
	|e_i\rangle=\sum_{n=0}^\infty\sum_{a^n} \langle a^n|e_i\rangle\,|a^n\rangle\quad.
\end{equation}
Now regard the set ${\cal E}:=\{|e_i\rangle\}_i$ itself as an alphabet, whose letters are the vectors $|e_i\rangle$, distributed by $q_i$. Then there is a Huffman code mapping each $|e_i\rangle$ to a unique binary codeword 
$|c(e_i)\rangle=|(c_1\cdots c_{l_i})(e_i)\rangle$,
which is a string of length 
$l_i=-\log q_i$,
taken from the binary basis alphabet ${\cal B}_C=\{|0\rangle,|1\rangle\}$.
Again, the above length is \emph{ideal}. In real life the Huffman code choses a codeword with an integer length next above $(-\log q_i)$.
Every eigenvector $|e_i\rangle$ of $\sigma$ is mapped to a string $|c(e_i)\rangle$ with
$\langle c(e_i)|c(e_j)\rangle=\delta_{ij}$,
and
\begin{equation}
	\widehat L_C|c(e_i)\rangle=l_i\,|c(e_i)\rangle
\end{equation}
by the encoder
\begin{equation}
	C_\Gamma:=\sum_i |c(e_i)\rangle\langle e_i|\quad.
\end{equation}
The encoder is a isometric operator on the \emph{source message space}
\begin{equation}
	{\cal M}_\Gamma=\text{Span}(\Gamma)\quad,
\end{equation}
i.e.
$C_\Gamma^\dagger C_\Gamma={\mathbbm1}_{{\cal M}_\Gamma}$.
Message components outside ${\cal M}_\Gamma$ are translated to the code space in the following way. Let
\begin{equation}
	\Pi_\Gamma:=\sum_i |e_i\rangle\langle e_i|
\end{equation}
be the projector onto the subspace ${\cal M}_\Gamma$ and $T:{\cal M}_{\cal Q}\rightarrow{\cal M}_C$ be a translator from the source alphabet ${\cal Q}$ to the code alphabet ${\cal Q}_C$ (see section~\ref{translation}), that fulfills
\begin{equation}
	\langle c(e_i)|T|\psi\rangle=0
	\quad\forall \psi\in {\cal M}_\Gamma^\perp,\forall i\quad,
\end{equation}
with ${\cal M}_\Gamma^\perp$ being the subspace orthogonal to ${\cal M}_\Gamma$. Hence the operator
\begin{equation}
	T_{\neg \Gamma}:=\big({\mathbbm 1}-\Pi_\Gamma\big)\,T\,
	\big({\mathbbm 1}-\Pi_\Gamma\big)
\end{equation} 
translates only message components outside the subspace ${\cal M}_\Gamma$ into code messages being orthogonal to any of the $|c(e_i)\rangle$, i.e.
$C_\Gamma^\dagger T_{\neg \Gamma}=T_{\neg \Gamma}^\dagger C_\Gamma
	=\widehat 0$.
That way the total encoder
\begin{equation}
	C:= C_\Gamma+T_{\neg \Gamma}
\end{equation}
is an isometric encoder from the source space ${\cal M}_{\cal Q}$ to the code space ${\cal M}_C$.
The encoded length is observed by
\begin{eqnarray}
	\widehat L_c&=&C^\dagger\,\widehat L_C\,C
	=C_\Gamma^\dagger\,\widehat L_C\,C_\Gamma
	+T_{\neg\Gamma}^\dagger\,\widehat L_C\,T_{\neg\Gamma}\\
	&=&\sum_i l_i\,|e_i\rangle\langle e_i|
	+R\,({\mathbbm1}-\Pi_\Gamma)\,\widehat L\,({\mathbbm1}-\Pi_\Gamma),
\end{eqnarray}
where $R$ is the rate of the translation code $T$ fulfilling
\begin{equation}
	R\geq\log(\dim{\cal H}_{\cal Q})\quad.
\end{equation}
The encoded information is observed by
\begin{eqnarray}
	\widehat I_c&=&\widehat L_c=\sum_i l_i\,|e_i\rangle\langle e_i|
	+\frac{R}{\log(\dim{\cal H}_{\cal Q})}
	\,\widehat I_{\neg\Gamma}\quad,
\end{eqnarray}
where
\begin{equation}\label{i_neg_gamma}
	\widehat I_{\neg\Gamma}:=({\mathbbm1}-\Pi_\Gamma)\widehat I
	({\mathbbm1}-\Pi_\Gamma)
\end{equation}
observes the information content of components outside ${\cal M}_\Gamma$.
Any \emph{a priori} message $|\varphi\rangle\in\Gamma$ from Alice is encoded into a superposition of Huffman strings of distinct lengths. 
Alice's entire source message ensemble $\sigma$, given by~(\ref{sigma}) obtains an encoded length of
$L_c(\sigma)=\sum_i q_i\,l_i$,
so the encoded ensemble information reads
\begin{equation}
	 I_c(\sigma)=\sum_i q_i\,l_i\quad.
\end{equation}
For an ideal Huffman code providing $l_i=-\log q_i$ the above value is minimized to
\begin{equation}
	 I_c(\sigma)=-\sum_i q_i\,\log q_i=S(\sigma)\quad.
\end{equation}
In the case of canonical messages $\sigma=\rho^{\otimes N}$ the encoded information reads
\begin{equation}
	 I_c(\rho^{\otimes N})=S(\rho^{\otimes N})=N\,S(\rho)\quad,
\end{equation}
hence optimal compression is achieved in any case.

\subsection{Core quantum information content}

In analogy to section~\ref{core_inf_cont} one may define an observable \emph{core information content} respecting a source ensemble $\sigma$, given by 
\begin{equation}
	\sigma=\sum_{\varphi\in\Gamma}p(\varphi)\,|\varphi\rangle\langle\varphi|
	=\sum_i q_i\,|e_i\rangle\langle e_i|\quad.
\end{equation}
For an ideal code the rate $R$ of the translation part fulfills
$R=\log(\dim{\cal H}_{\cal Q})$,
whereas the lengths of the compression part fulfill
$l_i=-\log q_i$.
Hence the \emph{core information content} can be defined as
\begin{equation}
	\widehat I_0:=-\log\sigma+\widehat I_{\neg\Gamma}\quad,
\end{equation}
where $\widehat I_{\neg\Gamma}$, given by~(\ref{i_neg_gamma}), measures the uncompressed information content outside ${\cal M}_\Gamma$, and $(-\log\sigma)$ measures the compressed information content inside ${\cal M}_\Gamma$.
The \emph{core information content} of a general message $\rho\in{\cal S}({\cal M}_{\cal Q})$ is then defined by
\begin{equation}
	I_0(\rho):=\text{Tr}\{\rho\,\widehat I_0\}\quad.
\end{equation}
The above value indicates the number of qbits being engaged on the average by communicating $\rho$ over a lossless channel that is fully optimized respecting the ensemble $\sigma$. 
For example, the core information of each \emph{a priori} message $|\varphi\rangle\in\Gamma$ that Alice sends, is given by
\begin{equation}
	I_0(\varphi)=-\langle\varphi|\log\sigma|\varphi\rangle
	=-\sum_i\log q_i\,|\langle e_i|\varphi\rangle|^2.
\end{equation}
Any other message $|\psi\rangle\in{\cal M}_{\cal Q}$ may also be sent without loss of information, but its compression is not optimized and might be poor, indicated by large values of $I_0(\psi)$.
The core information content of the source matrix itself equals its von Neumann entropy
\begin{equation}
	I_0(\sigma)=\sum_{\varphi\in\Gamma}p(\varphi)\,I_0(\varphi)
	=-\text{Tr}\{\sigma\,\log\sigma\}
	=S(\sigma).
\end{equation}
In this very sense the von Neumann entropy is the \emph{core quantum information} contained in a message matrix $\sigma$. For any given $\sigma$ Alice can design a lossless quantum code that minimizes the effort of communicating all \emph{a priori} message ensembles being equivalent to $\sigma$. The core information is a quantum mechanical observable that yields the \emph{number of qbits} that would be engaged if the message were communicated using a lossless compression code optimized for $\sigma$. The average core information of any message ensemble equivalent to $\sigma$ equals its von Neumann entropy. This confirms the meaning that is commonly assigned to the von Neumann entropy and puts it on a solid ground.

\section{Summary}

Within the framework of many-letter theory, a general characterization of quantum codes using the Kraus representation of completely positive maps has been given. An observable has been constructed measuring the \emph{raw quantum information content} of a particular message, where the unit of its value has been given the name ``1 qbit''. This type of quantum information content is merely related to the effort it takes to communicate a particular quantum message. It is not based on statistical properties of a message ensemble. Compression codes are defined  by their property of reducing the quantum information content of a given message ensemble. A general form of translation codes has been given that translate between two alphabets without loss of information. It is shown that these codes, as expected, are never compressive. The formalism has then been applied to the Schumacher coding scheme, which is only defined on a special type of messages, so-called \emph{canonical messages}, to see that the expected quantum information content per letter, represented by the introduced observable, can be reduced to the von-Neumann entropy, according to the known result. The Schumacher coding scheme has then been extended to a more general type of messages, so-called \emph{grand canonical messages}. However, as the Schumacher code can only be applied to messages of this type, the von Neumann entropy has not yet obtained its fundamental meaning. This has been changed by constructing a lossless coding scheme for \emph{all} messages providing optimal compression and perfect retrieval of the original data. The given coding scheme exploits the features of many-letter spaces and cannot be implemented in standard block Hilbert spaces. Motivated by the concept of lossless compression, an observable is constructed measuring the \emph{core information content} of a particular message with respect to a given \emph{a priori} message ensemble. The expectation value of the \emph{a priori} message ensemble itself equals its von Neumann entropy. Hence, in the context of lossless compression, the von Neumann entropy can be interpreted as the expected core quantum information content of a message ensemble that remains when any redundancy due to statistical predictability has been removed. This confirms the commonly assigned meaning of the von Neumann entropy.

\section{Acknowledgements}

I would like to thank Jens Eisert, Timo Felbinger, Alexander Albus, and Shash Virmani for fruitful and intensive discussions about the topic of this paper.


\end{multicols}

\end{document}